\documentclass{article}
\usepackage{mathrsfs}
\usepackage{amsmath,amsthm,amsfonts,amscd,eucal,latexsym,amssymb}
\usepackage{hyperref}

\usepackage[english]{babel}
 \newtheorem{thm}{Theorem}[section]
 
 \newtheorem{lem}[thm]{Lemma}
 \newtheorem{prop}[thm]{Proposition}
 \theoremstyle{definition}
 \newtheorem{defn}[thm]{Definition}
 \theoremstyle{remark}

 \numberwithin{equation}{section}
\begin{document}

\par 
\bigskip 
\par 
\rm 
 
\par 
\bigskip 
\LARGE 
\noindent 
{\bf Hadamard states from null infinity} 
\bigskip 
\par 
\rm 
\normalsize 

\large
\noindent {\bf Claudio Dappiaggi$^{1,2,a}$},\\
\par
\small
\noindent $^1$ 
Dipartimento di Fisica, Universit\`a degli Studi di Pavia,
Via Bassi, 6, 27100 Pavia, Italy.\smallskip

\noindent$^2$  Istituto Nazionale di Fisica Nucleare - Sezione di Pavia,
Via Bassi, 6, 27100 Pavia, Italy.\smallskip

\noindent $^a$  claudio.dappiaggi@unipv.it,\\
\normalsize

\par 
 
\rm\normalsize 
\noindent {\small Version of \today}

\rm\normalsize 
 
 
\par 
\bigskip 

\noindent 
\small 
{\bf Abstract}. 
Free field theories on a four dimensional, globally hyperbolic spacetime, whose dynamics is ruled by a Green hyperbolic partial differential operator, can be quantized following the algebraic approach. It consists of a two-step procedure: In the first part one identifies the observables of the underlying physical system collecting them in a $*$-algebra which encodes their relational and structural properties. In the second step one must identify a quantum state, that is a positive, normalized linear functional on the $*$-algebra out of which one recovers the interpretation proper of quantum mechanical theories via the so-called Gelfand-Naimark-Segal theorem. In between the plethora of possible states, only few of them are considered physically acceptable and they are all characterized by the so-called Hadamard condition, a constraint on the singular structure of the associated two-point function. Goal of this paper is to outline a construction scheme for these states which can be applied whenever the underlying background possesses a null (conformal) boundary. We discuss in particular the examples of a real, massless conformally coupled scalar field and of linearized gravity on a globally hyperbolic and asymptotically flat spacetime.    
\normalsize
\bigskip 

\section{Introduction}

Quantum field theory on curved backgrounds has witnessed a period of renaissance in the past twenty years. From a physical perspective, cosmology is acquiring an everyday greater relevance, mostly on account of the expected, upcoming plethora of experimental data. The models with which these should be compared are often aimed at the description of the dynamics of the evolution of the early Universe and are mainly based on fields living on homogeneous and isotropic manifolds. In this framework quantum effects are expected to play a key role, for example in shaping the fluctuations of the cosmic microwave background. 

Studying quantum field theory beyond Minkowski spacetime has shifted from a purely academic question to a concrete necessity and, thus, it has been accompanied contemporary by an increasing interest in developing its structural, foundational and mathematical 
aspects. In this respect a framework appears to claim the lion's share of interest: algebraic quantum field theory. In a few words, this is a mathematically rigorous approach based on two key steps. In the first, one codifies the observables of the physical system under investigation into a $*$-algebra $\mathcal{A}$ which encompasses the information about the dynamics, locality and causality. In the 
second, instead, one identifies a so-called algebraic state, namely a positive, normalized linear functional $\omega:\mathcal{A}\to\mathbb{C}$. Via the Gelfand-Naimark-Segal (GNS) theorem, one associates to the pair $(\mathcal{A},\omega)$ a triplet of data which in particular identifies a Hilbert space, as well as both a representation of $\mathcal{A}$ in terms of linear 
operators thereon and a unit norm, cyclic vector. In other words one recovers the standard interpretation of quantum mechanical theories. Much has been written about the algebraic approach, especially on curved backgrounds, and we refer an interested reader to the several recent reviews, {\it e.g.}, \cite{Benini:2013fia, Fredenhagen:2014lda}. 

We will be interested, instead, in a more specific problem. For any given algebra $\mathcal{A}$ one can easily construct several algebraic states, but one can hardly call all of them physically relevant. The reasons are manifold and often related to pathological behaviours such as for example the occurrence of divergences in the quantum fluctuations of observables or the impossibility of constructing Wick polynomials, which are the basis to deal with interactions at a perturbative level. Especially for free field theories on globally hyperbolic spacetimes this problem has been thoroughly discussed since already the seventies and by now it is almost unanimously accepted that a state, to be called physical, must be of {\em Hadamard form}. Originally it was formulated as a condition on the form in each geodesically convex neighbourhood of the manifold of the integral kernel of the two-point function. Verifying it in concrete cases is rather complicated if not outright impossible and, for this reason, for many years only few examples of Hadamard states were known, {\it e.g.}, the Poincar\'e vacuum on Minkowski spacetime and the Bunch-Davies state for a scalar field on de Sitter spacetime. Although generic existence results were known, a cornerstone in our understanding of Hadamard states came from the papers of Radzikowski \cite{Radzikowski:1996pa, Radzikowski:1996ei}. He showed that the Hadamard condition is fully equivalent to assigning a precise form to the wavefront set of the bi-distribution associated to the two-point correlation function of a quasi-free/Gaussian state. Despite the necessity of using microlocal analysis, controlling explicitly the wavefront set turns out to be much easier in concrete scenarios and hence, starting from  \cite{Radzikowski:1996pa, Radzikowski:1996ei}, many construction schemes for Hadamard states were devised. In this paper we review a particular one which originates from one of the possible transpositions in the algebraic framework of the often mentioned holographic principle. Often called also bulk-to-boundary correspondence, this constructive scheme was first discussed in \cite{Dappiaggi:2005ci} and it is devised to work for free field theories on four dimensional globally hyperbolic spacetimes possessing a null (conformal) boundary. The main idea is based on the observation that the generators of the observables of a free field theory are in correspondence with smooth and spacelike compact solutions of the underlying equations of motion. On account of the properties of the Green operators associated to the underlying dynamics, each of these solutions propagates to the null boundary identifying thereon a smooth function. Hence, by constructing a suitable $*$-algebra of functions on the null boundary, one can identify the algebra of observables of a free field theory with a $*$-subalgebra of the boundary one. More importantly this entails that each algebraic state on the boundary identifies a counterpart for the free field theory living on the bulk spacetime. The net advantage of this procedure is twofold: On the one hand the boundary usually possesses an infinite dimensional symmetry group, which, exactly as the Poincar\'e group in Minkowski spacetime, allows to identify a distinguished state thereon. On the other hand, the theorem of propagation of singularities in combination with a control of the wavefront set of the Green operators allows us to prove that such distinguished, boundary state induces a bulk counterpart of Hadamard form.

Up to now this procedure has been applied successfully in several different contexts and for different theories ranging from scalar fields on cosmological spacetimes \cite{Dappiaggi:2007mx, Dappiaggi:2008dk} and on black hole spacetimes \cite{Brum:2014nea, Dappiaggi:2008dk}, to Dirac fields on cosmological and asymptotically flat spacetimes \cite{Dappiaggi:2010gt}, to free electromagnetism \cite{Dappiaggi:2011cj}. It is noteworthy that this procedure allows also for the identification of local ground states \cite{Dappiaggi:2010iq} and it is suitable to be translated in the language of pseudodifferential calculus, as shown recently in these interesting papers \cite{Gerard:2014hla, Gerard:2012wb, Gerard:2014wb}. 

Since we cannot go into the details of all these results, we decided to focus our attention on two special but instructive applications of the bulk-to-boundary correspondence. The first discusses the procedure for a real, massless and conformally coupled scalar field on asymptotically flat spacetimes \cite{Dappiaggi:2005ci}. The second instead aims at reviewing the most recent application of this construction and at showing the additional complications arising for linear gauge theories, namely we shall discuss linearized gravity following \cite{Benini:2014rya}.

The paper is organized as follows: In Section \ref{2} we outline the classical dynamics of a real scalar field and the construction of the algebra of fields. In Section \ref{3}, first we introduce the class of asymptotically flat and globally hyperbolic, four dimensional spacetimes and, subsequently, we discuss the bulk-to-boundary correspondence and particularly the construction of a Hadamard state for a massless, conformally coupled scalar field. In Section \ref{4} we focus our attention on linearized gravity and we show how to construct an algebra of fields while dealing with gauge freedom. To conclude, in Section \ref{5}, we repeat briefly the construction of Hadamard states starting from future null infinity, focusing mainly on the additional problems brought by gauge invariance. On account of the lack of space, we plan to avoid giving detailed proofs of all mathematical statements, referring each time instead to the relevant literature. 

\section{Scalar Field Theory}\label{2}
In this section, we shall outline the quantization within the algebraic framework of the simplest example of field theory. First of all we need to specify the class of backgrounds which we consider as admissible and which, henceforth, are referred to as spacetimes. We allow only four dimensional, connected, smooth manifolds $M$ endowed with a smooth Lorentzian metric $g$ of signature $(-,+,+,+)$. Furthermore we require $(M,g)$ to be globally hyperbolic, that is $M$ possesses a Cauchy surface $\Sigma$, a closed achronal subset of $M$ whose domain of dependence coincides with the whole spacetime -- for more details, refer to \cite[Ch. 8]{Wald}. The existence of a Cauchy surface leads to several noteworthy consequences. In between them, we stress that the property of being globally hyperbolic entails that $M$ is isometric to the Cartesian product $\mathbb{R}\times\Sigma$ and thereon there exists a coordinate system such that the line element reads 
$$ds^2=-\beta dt^2+h_t,$$
where $t:\mathbb{R}\times\Sigma\to\mathbb{R}$ is the projection on the first factor, while $\beta$ is a smooth and strictly positive scalar function on $\mathbb{R}\times\Sigma$. Furthermore, for all values of $t$, $\{t\}\times\Sigma$ is a $3$-dimensional smooth, spacelike, Cauchy surface in $M$ and $t\mapsto h_t$ is a one-parameter family of smooth Riemannian metrics -- see \cite{Bar:2007zz} and references therein.

Besides a globally hyperbolic spacetime $(M,g)$, we consider a real scalar field $\phi:M\to\mathbb{R}$ whose dynamics is ruled by the Klein-Gordon equation
\begin{equation}\label{KG}
P\phi=\left(\Box-m^2-\xi R\right)\phi=0,
\end{equation}
were $m\geq 0$ is the mass, $R$ is scalar curvature built out of $g$, while $\xi\in\mathbb{R}$ is a coupling constant. While all values of $\xi$ are admissible and, from a structural point of view, they are all equivalent, two cases are often considered in the literature: $\xi=0$, also known as {\em minimal coupling} or $\xi=\frac{1}{6}$, the {\em conformal coupling}. In this paper we will be interested mainly in this last option, moreover with a vanishing mass. Regardless of the value of $m$ and $\xi$, the operator $P$ is a special case of a Green hyperbolic partial differential operator -- refer to \cite{Bar:2013}. Hence, the smooth solutions of \eqref{KG} can be constructed in terms of a Cauchy problem, for which smooth initial data are assigned on any but fixed Cauchy surface $\{t\}\times\Sigma$. Yet, this approach breaks manifestly covariance and, especially from the perspective of the algebraic quantization scheme, it is more appropriate to adopt a different approach, namely that of Green functions. To this end we introduce a notable class of functions:
\begin{defn}\label{tc}
We call $C^\infty_{tc}(M)$, the collection of all {\em timelike compact} functions, that is $f\in C^\infty_{tc}(M)$ if $f$ is a smooth function such that, for all $p\in M$, $\textrm{supp}(f)\cap J^\pm(p)$ is either empty or compact. Here $J^\pm(p)$ stands for the causal future (+) and past (-) of $p$.
\end{defn}
Following \cite{Bar:2013, Sanders:2012ac} and generalizing slightly the content of \cite{Bar:2009zzb, Bar:2007zz}, we introduce
\begin{defn}\label{Green}
We call {\em retarded} (+) and {\em advanced Green operators} (-) associated to the Klein-Gordon operator $P$, two linear maps $E^\pm: C^\infty_{tc}(M)\to C^\infty(M)$ such that, for every $f\in C^\infty_{tc}(M)$,
$$\left(P\circ E^\pm\right)(f)=f=\left(E^\pm\circ P\right)(f),$$
and $\textrm{supp}\left(E^\pm(f)\right)\subseteq J^\pm(\textrm{supp}(f))$. The map $E=E^--E^+$ is called {\em causal propagator}.
\end{defn}
If we introduce the space of all smooth solutions of \eqref{KG},
\begin{equation}\label{sol}
\mathcal{S}(M)\doteq\{\phi\in C^\infty(M)\;|\;P\phi=0\},
\end{equation}
advanced and retarded Green operators can be used to prove two important properties of \eqref{KG} which we only recollect here:
\begin{itemize}
\item There exists an isomorphism of topological vector spaces between $\mathcal{S}(M)$ and $\frac{C^\infty_{tc}(M)}{P[C^\infty_{tc}(M)]}$ which is realized by $E$. In other words, for every $\phi\in\mathcal{S}(M)$, there exists $[f]\in\frac{C^\infty_{tc}(M)}{P[C^\infty_{tc}(M)]}$ such that $\phi=E(f)$ regardless of the chosen representative in $[f]$.
\item Let $\mathcal{S}_{sc}(M)\subset \mathcal{S}(M)$ be the vector subspace of the {\em spacelike compact}, smooth solutions $\phi$ to \eqref{KG}, that is $\textrm{supp}(\phi)\cap \left(\{t\}\times\Sigma)\right)$ is compact for all values of $t$. Still via the causal propagator, $\mathcal{S}_{sc}(M)$ is isomorphic to $\frac{C^\infty_0(M)}{P[C^\infty_0(M)]}$ where $C^\infty_0(M)$ is the collection of smooth and compactly supported functions on $M$. Most notably $\mathcal{S}_{sc}(M)$ is a symplectic space if endowed with the following weakly non-degenerate symplectic form $\sigma:\mathcal{S}_{sc}(M)\times\mathcal{S}_{sc}(M)\to\mathbb{R}$ 
\begin{equation}\label{sympl}
\sigma(\phi,\phi^\prime)=E([f],[f^\prime])\doteq\int\limits_M d\mu_g E(f)f^\prime,
\end{equation}
where $d\mu_g$ is the metric-induced volume form, while $[f],[f^\prime]\in \frac{C^\infty_0(M)}{P[C^\infty_0(M)]}$ are such that $\phi^{(\prime)}=E(f^{(\prime)})$.
\end{itemize}

Having under control the space of all smooth solutions of \eqref{KG}, we can introduce the notion of observable for a real scalar field. Notice that, at this stage, we are still working at a purely classical level. The underlying paradigm is that an observable is nothing but an assignment of a real number to any configuration of a physical system, done in a way which is compatible with the underlying dynamics. At a mathematical level, this heuristic statement can be translated as follows: Let us consider {\em off-shell/kinematical} configurations, namely all $\phi\in C^\infty(M)$ and, for any $f\in C^\infty_0(M)$, we define the linear functional
\begin{equation}\label{functional}
F_f:C^\infty(M)\to\mathbb{R},\quad\phi\mapsto F_f(\phi)\doteq\int\limits_M d\mu_g\;\phi(x)f(x).
\end{equation}
The map $F_f$ plays the role of a classical linear observable for the kinematical configurations of a Klein-Gordon field and, from standard results in functional analysis, we can also infer that the assignment $f\mapsto F_f$ is injective and that the collection of all functionals, built in this way, is separating. This entails that, for every $\phi,\phi^\prime\in C^\infty(M)$, we can find at least one $f\in C^\infty_0(M)$ such that $F_f(\phi)\neq F_f(\phi^\prime)$. 

In order to codify in this construction the information of the dynamics, it suffices to restrict the kinematical configurations to the dynamically allowed, namely to $\mathcal{S}(M)\subset C^\infty(M)$. In other words we consider now the assignment $f\in C^\infty_0(M)\mapsto F_f:\mathcal{S}(M)\to\mathbb{R}$. While the property of being a separating set is left untouched by the restriction of the configurations allowed, injectivity is no longer valid. As a matter of fact one can prove that $F_f(\phi)=0$ for all $\phi\in\mathcal{S}(M)$ if and only if there exists $h\in C^\infty_0(M)$ such that $f=P(h)$. For this reason we identify those linear functionals $F_f$ and $F_{f^\prime}$ on $\mathcal{S}(M)$ such that $f-f^\prime\in P[C^\infty_0(M)]$. To summarize, the space of {\em linear classical observables} is 
\begin{equation}\label{obs}
\mathcal{E}^{obs}(M)\doteq\left\{\left. F_{[f]}:\mathcal{S}(M)\to\mathbb{R}\;\right|\; [f]\in\frac{C^\infty_0(M)}{P[C^\infty_0(M)]}\right\},
\end{equation} 
where $F_{[f]}(\phi)=F_f(\phi)$, the right hand side being defined in \eqref{functional}. On the one hand notice that $\mathcal{E}^{obs}(M)$ is isomorphic to the labeling space $\frac{C^\infty_0(M)}{P[C^\infty_0(M)]}$ and thus it comes with a symplectic form \eqref{sympl}. On the other hand we remark that the choice of observables is far from being unique. Our guiding principles have been essentially three: We want $\mathcal{E}^{obs}(M)$ to separate all dynamical configurations, to lack any redundant observable and to be a symplectic space. While in the analysis of a real scalar field, our approach might look as an overkill, which ultimately yields already well-known results, we stress that the paradigm that we used is very effective as soon as we deal with gauge theories -- see for example \cite{Benini}.

Having under control the dynamics of a real scalar field and having chosen a linear space of classical observables is the starting point of the quantization scheme, that we follow. As mentioned in the introduction the algebraic scheme of quantization is based on two steps, the first of which consists of regrouping all observables into a suitable $*$-algebra. Starting from \eqref{obs}, 

\begin{defn}\label{F(M)}
We call {\bf algebra of fields} for a real scalar field, whose dynamics is ruled by \eqref{KG}, the quotient $\mathcal{F}(M)=\frac{\mathcal{T}(M)}{\mathcal{I}(M)}$. Here 
$$\mathcal{T}(M)\doteq \bigoplus_{n=0}^\infty\mathcal{E}^{obs}(M;\mathbb{C})^{\otimes n},$$
is the universal tensor algebra endowed with complex conjugation as $*$-operation, where $\mathcal{E}^{obs}(M;\mathbb{C})=\mathcal{E}^{obs}(M)\otimes\mathbb{C}$ while $\mathcal{E}^{obs}(M;\mathbb{C})^{\otimes 0}\doteq\mathbb{C}$. $\mathcal{I}(M)$ is the $*$-ideal generated by the {\em canonical commutation relations} $[f]\otimes [f^\prime]-[f^\prime]\otimes [f]-iE\left([f],[f^\prime]\right)\mathbb{I}$, where $\mathbb{I}$ is the identity in $\mathcal{T}(M)$.
\end{defn}

Notice that, in the definition, we have implicitly used the identification between classical observables and the labeling space $\frac{C^\infty_0(M)}{P[C^\infty_0(M)]}$ while $E\left([f],[f^\prime]\right)$ is the symplectic form \eqref{sympl}. Although much has been written in the literature about $\mathcal{F}(M)$, we want to recall here two of its most important properties, the proof of which can be found for example in \cite{Benini:2014rya}.

\begin{prop}\label{properties}
The algebra of fields $\mathcal{F}(M)$
\begin{enumerate}
\item is {\bf causal}, that is elements of $\mathcal{F}(M)$ localized in causally disjoint regions commute,
\item fulfills the {\bf time-slice axiom}. Let $\mathcal{O}\subset M$ be a globally hyperbolic open neighborhood 
of a spacelike Cauchy surface $\Sigma$, that is $\mathcal{O}$ contains all causal curves for $(M,g)$ whose endpoints lie in $\mathcal{O}$. Let $\mathcal{F}(\mathcal{O})$ be the algebra of fields for a real scalar field in $(\mathcal{O},\left.g\right|_{\mathcal{O}})$. Then $\mathcal{F}(\mathcal{O})$ is $*$-isomorphic to $\mathcal{F}(M)$.
\end{enumerate}
\end{prop}

We have completed the first part of the algebraic quantization procedure and, thus, we are left with the second and last one. From a more general point of view, given any unital $*$-algebra $\mathcal{A}$, we call {\em algebraic state} a positive and normalized linear functional $\omega:\mathcal{A}\to\mathbb{C}$ such that 
$$\omega(\mathbb{I})=1,\qquad\omega(a^*a)\geq 0,\;\forall a\in\mathcal{A}$$
where $\mathbb{I}$ is the unit of $\mathcal{A}$. States are relevant since, in combination with $\mathcal{A}$, they allow for recovering the probabilistic interpretation proper of quantum theories via the celebrated GNS theorem, which we recollect here -- see \cite[Ch. 14]{Moretti:2013cma}
\begin{thm}\label{GNS}
Let $\omega$ be a state on a unital $*$-algebra $\mathcal{A}$. 
There exists a dense subspace $\mathcal{D}$ of a Hilbert space $(\mathcal{H},(\cdot,\cdot))$, 
as well as a representation $\pi:\mathcal{A}\to\mathcal{L}(\mathcal{D})$ 
and a unit norm, cyclic vector $\Omega\in\mathcal{D}$, 
such that $\omega(\cdot)=(\Omega,\pi(\cdot)\Omega)$ and $\mathcal{D}=\pi(\mathcal{A})\Omega$.
The {\em GNS triple} $(\mathcal{D},\pi,\Omega)$ is uniquely determined up to unitary equivalence.
\end{thm}

We are interested especially in the case where the $*$-algebra is $\mathcal{F}(M)$. In this respect, constructing algebraic states is not a difficult operation and several examples are easily available. Yet, in between all of them, the most part cannot be considered physically acceptable and one of the biggest challenges in the algebraic approach to quantum field theory has been indeed to identify suitable criteria to single out states which yield a good physical behaviour. Especially when one cannot exploit Poincar\'e covariance to construct a (unique) vacuum state, {\it e.g.}, when the underlying background is curved, the question is of paramount importance. After long debates it is nowadays almost unanimously accepted that the answer consists of requiring that an algebraic state fulfills the so-called {\em Hadamard condition}. 

From a physical perspective it amounts to saying that a state is physically acceptable if all quantum fluctuations of the observables are finite and if its ultraviolet behaviour coincides with that of the Poincar\'e vacuum. The translation at a mathematical level is the following. As a starting point we notice that, since $\mathcal{F}(M)$ is the quotient between $\mathcal{T}(M)$ and $\mathcal{I}(M)$, assigning a state $\omega:\mathcal{F}(M)\to\mathbb{C}$ is equivalent to constructing all $n$-point correlation functions $\omega_n:\left(C^\infty_0(M)\right)^{\otimes n}\to\mathbb{C}$ fulfilling suitable constraints so to ensure positivity and compatibility both with the dynamics and with the canonical commutations relations. In between all states, most notable are the quasi-free/Gaussian ones, which are characterized by the property of being completely determined by the associated two-point function. With a slight abuse of notation, this is tantamount to choosing a suitable $\omega_2\in\mathcal{D}^\prime(M\times M)$, extending it via the formula:
$$\omega_{2n}(\lambda_1\otimes...\otimes\lambda_{2n})=\sum\limits_{\pi_{2n}\in S_{2n}}\prod\limits_{i=1}^{n}\omega_2(\lambda_{\pi_{2n}(2i-1)}\otimes\lambda_{\pi_{2n}(2i)}),\quad\omega_{2n+1}=0,$$
where $n\in\mathbb{N}$. Quasi-free states allow to formulate the Hadamard property as a constraint on the singular structure of the associated $\omega_2$:
\begin{defn}\label{Hadamard}
A quasi-free state $\omega:\mathcal{F}(M)\to\mathbb{C}$ is called {\bf Hadamard} if the associated two-point function $\omega_2\in\mathcal{D}^\prime(M\times M)$ is such that its wavefront set has the following form:
\begin{equation}\label{WF}
WF(\omega_2)=\left\{(x,y,k_x,-k_y)\in T^*{\!M}^2\setminus \{\mathbf 0\}\;|\;
(x,k_x)\sim(y,k_y),\;k_x\triangleright 0\right\}\,.
\end{equation}
Here, $(x,k_x)\sim(y,k_y)$ implies that there exists a null geodesic $c$ connecting $x$ to $y$ such
that $k_x$ is coparallel and cotangent to $c$ at $x$ and $k_y$ is the parallel transport of $k_x$ from $x$ to $y$ along $c$. Finally, $k_x\triangleright 0$ means that the covector $k_x$ is future-directed.
\end{defn}

Unfortunately Definition \ref{Hadamard} is not constructive and, thus, one might wonder how to build concretely examples of Hadamard states. For many years only abstract existence results or example like the Poincar\'e vacuum and the Bunch-Davies state respectively on Minkowski and on the de Sitter spacetime were known, but, in the last decade, many novel construction schemes for Hadamard states, applicable in different frameworks, were devised. In the next section we shall review one of them which is especially suited for those linear field theories built on globally hyperbolic spacetimes possessing a null (conformal) boundary.
\subsection{Hadamard states from null infinity - I}\label{3}

In this section, we outline a procedure to construct explicitly Hadamard states, sometimes known as bulk-to-boundary correspondence. We will not work in full generality and we will be interested only in a massless, conformally coupled real scalar field, that is the dynamics is ruled by \eqref{KG} with $m=0$ and $\xi=\frac{1}{6}$. It is important to remark once more that the procedure, that we outline can be and has been applied to a wider range of free fields and of curved backgrounds, as we have already mentioned in the introduction. Since a full account of all these results would require much more space than that at our disposal, we start by reviewing the first application of this construction method, as it appeared in \cite{Dappiaggi:2005ci, Moretti:2005ty, Moretti:2006ks}. 

In particular, as starting point, we focus our attention on a particular class of globally hyperbolic spacetimes which are distinguished since they possess an asymptotic behaviour along null directions which mimics that of Minkowski spacetime. Used extensively and successfully in the definition of black hole regions \cite{Wald}, the most general class of asymptotically flat spacetimes includes several important physical examples, such as for instance the Schwarzschild and the Kerr solutions to Einstein's equations. Here we  employ the definition of asymptotic flatness, as introduced by Friedrich in \cite{Friedrich:1986uq}. To wit, we consider an \textbf{asymptotically flat spacetime with future timelike infinity $i^+$}, \textit{i.e.}, a globally hyperbolic spacetime $(M, g)$, hereby called \emph{physical spacetime}, such that there exists a second globally hyperbolic spacetime $(\widetilde M, \widetilde g)$, called \emph{unphysical spacetime}, with a preferred point $i^+\in\widetilde M$, a diffeomorphism $\Phi : M \to \Phi(M) \subset \widetilde M$ and a function $\Xi : \Phi(M) \to (0, \infty)$ so that $\Phi^\ast(\Xi^{-2}\widetilde g) = g$.
Furthermore, the following requirements ought to be satisfied:
\begin{itemize}
  \item[a)]
    If we call $J_{\widetilde M}^-(i^+)$ the causal past of $i^+$, this is a closed set such that $\Phi(M) = J_{\widetilde M}^-(i^+) \setminus \partial J_{\widetilde M}^-(i^+)$ and we have $\partial M = \partial J_{\widetilde M}^-(i^+) = \mathscr{I}^+ \cup \{i^+\}$, where $\mathscr{I}^+$ is called \emph{future null infinity}.
  \item[b)]
    $\Xi$ can be extended to a smooth function on the whole $\widetilde M$ and it vanishes on $\mathscr{I}^+ \cup \{i^+\}$.
    Furthermore, $d \Xi \neq 0$ on $\mathscr{I}^+$ while $d \Xi = 0$ on $i^+$ and $\widetilde \nabla_\mu \widetilde \nabla_\nu \, \Xi = -2 \, \widetilde g_{\mu\nu}$ at $i^+$.
  \item[c)]
    Introducing $n^\mu \doteq \widetilde \nabla^\mu \Xi$, there exists a smooth and positive function $\xi$ supported at least in a neighbourhood of $\mathscr{I}^+$ such that $\widetilde \nabla_\mu (\xi^4 n^\mu) = 0$ on $\mathscr{I}^+$ and the integral curves of $\xi^{-1}n$ are complete on future null infinity.
\end{itemize}
Notice that we shall henceforth identify $M$ with $\Phi(M)$. Here $\widetilde \nabla$ is the Levi-Civita connection built out of $\widetilde g$. Notice that, in the above definition, future timelike infinity plays a distinguished role, contrary to what happens in the more traditional definition of asymptotically flat spacetimes where $i^+$ is replaced by $i_0$, spatial infinity -- see for example \cite[Section 11]{Wald}. The reason for our choice is motivated by physics: We are interested in working with the algebra $\mathcal{F}(M)$ of Definition \ref{F(M)} which is constructed out of $E$, the causal propagator associated to the operator $P$ as in \eqref{KG} with $m=0$ and $\xi=\frac{1}{6}$. This entails in particular, that, for any smooth and compactly supported function $f$, its image under the action of the causal propagator is supported in the causal future and past of $\textrm{supp}(f)$. Therefore it will be important in our investigation that future timelike infinity is actually part of the unphysical spacetime, so to be able to control the behaviour of $E(f)$ thereon. Such requirement can be relaxed particularly if one is interested in studying field theories on manifolds like the Kruskal extension of Schwarzschild where $i^+$ cannot be made part of the unphysical spacetime. The price to pay in this case is the necessity to make sure that any solution of the classical dynamics falls off sufficiently fast as it approaches future timelike infinity. This line of reasoning has been pursued in \cite{Dappiaggi:2009fx}, though we shall not follow it here since it relies heavily on the fact that a very specific manifold has been chosen. On the contrary we work with a large class of backgrounds. 

Before proceeding towards the construction of a Hadamard state for a massless, conformally coupled, real scalar field on a globally hyperbolic and asymptotically flat spacetime, we point out a few distinguished properties of future null infinity -- see \cite{Moretti:2005ty, Wald} and references therein:
\begin{itemize}
\item $\mathscr{I}^+$ is a three dimensional, null submanifold of $\widetilde M$, diffeomorphic to $\mathbb{R}\times\mathbb{S}^2$. Furthermore there exists an open neighbourhood $U$of $\mathscr{I}^+$ and a coordinate system $(u,\Xi,\theta,\varphi)$, called {\em Bondi chart}, such that $(\theta,\varphi)$ are the standard coordinates on the unit $2$-sphere, $u$ is an affine parameter along the null geodesic generating $\mathscr{I}^+$, while $\Xi$ is the conformal factor, promoted to coordinate. In this system $\mathscr{I}^+$ is the locus $\Xi=0$ and the line elements reads
$$\left. ds^2\right|_{\mathscr{I}^+}= -2dud\Xi+d\theta^2+\sin^2\theta d\varphi^2.$$
\item There exists a distinguished subgroup of $Diff(\mathscr{I}^+)$, called the {\em Bondi-Metzner-Sachs (BMS) group} which can be defined via its action on a Bondi frame: If $(z,\bar{z})$ are the complex coordinates built out of $(\theta,\varphi)$ via a stereographic projection,
\begin{equation}\label{BMS}
\left\{
\begin{array}{l}
u\mapsto u^\prime\doteq K_\Lambda(z,\bar{z})\left(u+\alpha(z,\bar{z})\right)\\
z\mapsto z^\prime = \frac{az+b}{cz+d},\quad ad-bc=1,\quad \textrm{and c.c.}
\end{array}\right.
\end{equation}
where $a,b,c,d\in\mathbb{C}$, $\alpha\in C^\infty(\mathbb{S}^2)$,
\begin{equation}\label{KLambda}
\Lambda=\left[\begin{array}{cc}
a & b \\
c & d 
\end{array}\right],\quad\textrm{and}\quad K_\Lambda(z,\bar{z})=\frac{1+|z|^2}{|az+b|^2+|cz+d|^2}.
\end{equation}
By direct inspection, it turns out that \eqref{BMS} identifies the semidirect product $SL(2,\mathbb{C})\ltimes C^\infty(\mathbb{S}^2)$. Notice that the BMS group coincides, moreover, with the group of asymptotic symmetries of the physical spacetime $(M,g)$ \cite{Geroch}.
\end{itemize}

\noindent We have all the ingredients to implement the bulk-to-boundary correspondence mentioned at the beginning of the section. The basic procedure is based on two key structures. First of all, one defines on top of $\mathscr{I}^+$ a symplectic space of functions and an associated $*$-algebra of fields. Secondly one looks for an injective $*$-homomorphism from $\mathcal{F}(M)$, the algebra of fields defined in the physical spacetime $(M,g)$, and the one on $\mathscr{I}^+$. As a byproduct of such an homomorphism, every algebraic state on the boundary induces via pull-back a bulk counterpart. The net advantage is the fact that, at a geometric level, future null infinity comes endowed with the BMS group, which is an infinite dimensional symmetry group, which, exactly as the Poincar\'e group in Minkowski spacetime, allows to identify a distinguished ground state. 

\noindent In order to implement this programme, let us define:
\begin{equation}\label{Scrifunc}
\mathcal{S}(\mathscr{I}^+)\doteq\left\{\psi\in C^\infty(\mathscr{I}^+)\;|\;\psi\,\textrm{and}\,\partial_u\psi\in L^2(\mathscr{I}^+,d\mu_{\mathscr{I}})\right\},
\end{equation}
where $d\mu_\mathscr{I}=\sin\theta dud\theta d\varphi$. This is a symplectic space if endowed with 
$$\sigma_\mathscr{I}:\mathcal{S}(\mathscr{I}^+)\times\mathcal{S}(\mathscr{I}^+)\!\to\!\mathbb{R},\,(\psi,\psi^\prime)\mapsto\sigma_\mathscr{I}(\psi,\psi^\prime)=\int\limits_\mathscr{I}d\mu_\mathscr{I}\left(\psi\partial_u\psi^\prime-\psi^\prime\partial_u\psi\right).$$
Following the same procedure used starting from $\mathcal{E}^{obs}(M)$ in \eqref{obs},
\begin{defn}\label{Scriscalar}
We call {\em $*$-algebra of fields on $\mathscr{I}^+$}, $\mathcal{F}(\mathscr{I}^+)=\frac{\mathcal{T}(\mathscr{I}^+)}{\mathcal{I}(\mathscr{I}^+)}$, where  
$$\mathcal{T}(\mathscr{I}^+)\doteq\bigoplus_{n=0}^\infty\mathcal{S}(\mathscr{I}^+;\mathbb{C})^{\otimes n}.$$
Here $\mathcal{S}(\mathscr{I}^+;\mathbb{C})=\mathcal{S}(\mathscr{I}^+)\otimes\mathbb{C}$ whereas $\mathcal{S}(\mathscr{I}^+;\mathbb{C})^{\otimes 0}\doteq\mathbb{C}$. $\mathcal{I}(\mathscr{I}^+)$ is the $*$-ideal generated by the relation $\psi\otimes\psi^\prime-\psi^\prime\otimes\psi-i\sigma_{\mathscr{I}}(\psi,\psi^\prime)\mathbb{I}$, where $\mathbb{I}$ is the identity in $\mathcal{T}(\mathscr{I}^+)$. The $*$-operation is complex conjugation.
\end{defn}

At this stage it becomes clear why we chose to work only with a massless, conformally coupled, real scalar field. Since we want to construct a $*$-homomorphism between $\mathcal{F}(M)$ and $\mathcal{F}(\mathscr{I}^+)$, compatibility between the canonical commutation relations in $\mathcal{I}(M)$ and those in $\mathcal{I}(\mathscr{I}^+)$ suggest that we should start from an injective symplectomorphism between $(\mathcal{E}^{obs}(M),\sigma)$ and $(\mathcal{S}(\mathscr{I}^+),\sigma_\mathscr{I})$. In order to relate an equivalence class of compactly supported functions in $M$ and a smooth function on $\mathscr{I}^+$, the procedure calls for propagating the former to null infinity via the causal propagator $E$ associated to the underlying dynamics. Yet, one needs to remember that $\mathscr{I}^+$ is a submanifold of the unphysical spacetime which is related to the physical one by a conformal transformation. In this respect it is a well-known fact that \eqref{KG} is not well-behaved under such map, leading to pathological behaviours at $\mathscr{I}^+$ if a mass term or a coupling different from $\xi=\frac{1}{6}$ is present. On the contrary, as the name conformal coupling suggests, a solution to the massless and conformally coupled Klein-Gordon on $(M,g)$ has the property of staying a solution of the same equation on $(\widetilde M,\widetilde g)$ up to a conformal rescaling. More precisely the following holds true -- see for example \cite{Dappiaggi:2005ci, Wald}
\begin{prop}\label{confresc}
Let $(M,g)$ be a globally hyperbolic and asymptotically flat spacetime, whose associated unphysical spacetime $(\widetilde M,\widetilde g)$, $\left.\widetilde g\right|_{M}=\Xi^2 g$, is also globally hyperbolic. Let $R$ and $\widetilde R$ be the Ricci scalars built out of $g$ and $\widetilde g$ respectively. If $\phi$ is a smooth and spacelike compact solution of $P\phi=(\Box_g-\frac{R}{6})\phi$ on $(M,g)$, $\widetilde\phi=\Xi^{-1}\phi$ is a solution of $\widehat P\widetilde\phi=\left(\Box_{\widetilde g}-\frac{\widetilde R}{6}\right)\widetilde\phi=0$ on $(M,\widetilde g)$. Furthermore if $f\in C^\infty_0(M)$ is such that $\phi=E(f)$, $E$ being the causal propagator of $P$, then $\widetilde\phi=\left.\widetilde E_{\widehat{P}}(\Xi^{-3}f)\right|_M$, $\widetilde E_{\widehat{P}}$ being the causal propagator of $\widehat P$.
\end{prop}

As a by-product of this last proposition, we have associated to every observable in the physical spacetime a spacelike compact smooth solution of the massless, conformally coupled Klein-Gordon equation on $(\widetilde M,\widetilde g)$. Due to the support properties of the causal propagator, every such solution can be restricted to $\mathscr{I}^+$, which is a smooth submanifold of $\widetilde M$. The relevant map is
\begin{equation}\label{proj}
\Upsilon:\mathcal{E}^{obs}(M)\to C^\infty(\mathscr{I}^+)\quad[f]\mapsto\left.E_{\widehat{P}}(\Xi^{-3}f)\right|_{\mathscr{I}^+}.
\end{equation}
As proven in \cite{Moretti:2005ty, Moretti:2006ks} the following key property holds true:
\begin{prop}
The application $\Upsilon$, constructed in \eqref{proj}, is an injective linear map from $\mathcal{E}^{obs}(M)$ to $\mathcal{S}(\mathscr{I}^+)$ which is, moreover, a symplectomorphism. In other words, for every $[f],[f^\prime]\in\mathcal{E}^{obs}(M)$
$$\sigma\left([f],[f^\prime]\right)=\sigma_\mathscr{I}\left(\Upsilon([f]),\Upsilon([f^\prime])\right),$$
where $\sigma$ and $\sigma_\mathscr{I}$ are respectively the bulk and the boundary symplectic forms.
\end{prop}

Notice that this proposition allows us to extend the action of the projection map $\Upsilon$ at a level of algebra of fields. On the one hand we have automatically built a map from $\mathcal{T}(M)$ to $\mathcal{T}(\mathscr{I}^+)$, since $\Upsilon$ projects classical linear observables to elements in $\mathcal{S}(\mathscr{I}^+)$ which are respectively the generating space for the bulk and for the boundary universal tensor algebra. On the other hand, since $\Upsilon$ preserves at the same time the symplectic form, such map is compatible with the ideal generated by the canonical commutation relations on $M$ and by $\sigma_\mathscr{I}$ on $\mathscr{I}^+$. Since the $*$-operation is complex conjugation both in the bulk and in the boundary and since $\Upsilon$ leaves it untouched, we have the following lemma:
\begin{lem}
There exists an injective $*$-homomorphism $\iota:\mathcal{F}(M)\to\mathcal{F}(\mathscr{I}^+)$ which is defined by its action on the generators, namely $\iota([\alpha])\doteq\Upsilon([\alpha])$ where $\Upsilon$ is the map \eqref{proj}.
\end{lem}

\noindent The most important consequence of this lemma is the following: Let $\omega_\mathscr{I}:\mathcal{F}(\Im^+)\to\mathbb{C}$ be a normalized, positive linear functional, then 
$$\omega\doteq\iota^*\omega_\mathscr{I}:\mathcal{F}(M)\to\mathbb{C},\quad a\mapsto\omega(a)\doteq(\iota^*\omega_\mathscr{I})(a)=\omega_\mathscr{I}(\iota(a)),\,\forall a\in\mathcal{F}(M),$$
is a state for the algebra of fields on $(M,g)$.

As a consequence we can focus our attention on constructing algebraic states directly on null infinity, studying only subsequently the properties of the bulk counterpart, obtained via pull-back. As mentioned before, the advantage is the presence of the infinite dimensional BMS group on $\mathscr{I}^+$. Hence the best course of action is to build a quasi-free/Gaussian state for $\mathcal{F}(\mathscr{I}^+)$ by looking for a BMS invariant two-point function on future null infinity. This problem has been discussed thoroughly in different publications \cite{Dappiaggi:2009fx, Moretti:2005ty, Moretti:2006ks} and we report here the main results:

\begin{thm}
Let $\omega_{2,\mathscr{I}}:\mathcal{S}(\mathscr{I}^+)\otimes\mathcal{S}(\mathscr{I}^+)\to\mathbb{C}$ be
$$\omega_{2,\mathscr{I}}(\psi\otimes\psi^\prime)=-\frac{1}{\pi}\lim\limits_{\epsilon\to 0}\int\limits_{\mathbb{R}^2\times\mathbb{S}^2}dudu^\prime d\mathbb{S}^2(\theta,\varphi)\frac{\psi(u,\theta,\varphi)\psi^\prime(u^\prime,\theta,\varphi)}{(u-u^\prime-i\epsilon)^2},$$
where $d\mathbb{S}^2(\theta,\varphi)$ is the standard measure on the unit $2$-sphere. Then the following holds true:
\begin{enumerate}
\item $\omega_{2,\mathscr{I}}$ defines a quasi-free state $\omega_{\mathscr{I}}$ for $\mathcal{F}(\Im^+)$. In its folium this is the unique BMS invariant state.
\item The state $\omega\doteq\iota^*\omega_{\mathscr{I}}:\mathcal{F}(M)\to\mathbb{C}$ is a quasi-free state for the algebra of fields in the bulk which is 
\begin{itemize}
\item of Hadamard form,
\item invariant under the action of all isometries of $(M,g)$.
\end{itemize}
\end{enumerate}
\end{thm}

Notice that, invariance under all bulk isometries implies that our construction, applied to Minkowski spacetime, yields the Poincar\'e vacuum. 

\section{Linearized Gravity}\label{4}
The bulk-to-boundary correspondence, described in the previous section for a massless, conformally coupled real scalar field can be applied to other free fields and to other globally hyperbolic spacetimes possessing a null boundary. Barring minor technical details, the procedure is always the same except when one deals with non interacting gauge theories, since one needs to control additionally the gauge fixing. There are two cases which are certainly of relevance at a physical level: free electromagnetism and linearized gravity. The first was discussed a couple of years ago in \cite{Dappiaggi:2011cj}, while the second was only analyzed last year and therefore we  review it here, pointing out in particular the additional difficulties compared to the scalar case. We will summarize mainly the results of \cite{Fewster:2012bj} and of \cite{Benini:2014rya}. As in the previous sections, for the lack of space we prefer to avoid giving the proofs of our statements, each time referring a reader to the relevant literature. 

We consider still an arbitrary, globally hyperbolic and asymptotically flat spacetime $(M,g)$ with the additional constraint that the Ricci tensor vanishes, $Ric(g)=0$. In other words $(M,g)$ is a solution of the Einstein vacuum equations. On top of $(M,g)$ we consider a smooth symmetric $2$-tensor $h\in\Gamma(S^2T^*M)$ where $S^2T^*M\doteq T^*M\otimes_s T^*M$, the subscript $s$ standing for symmetrization. Dynamics is ruled by the linearized Einstein's equations:
\begin{equation}\label{lindyn}
\left(Kh\right)_{ab}\doteq-\frac{g_{ab}}{2}\left(\nabla^c\nabla^d h_{cd}-\Box\textrm{Tr}(h)\right)-\Box\frac{h_{ab}}{2}-\frac{1}{2}\nabla_a\nabla_b\textrm{Tr}(h)+\nabla^c\nabla_{(a}h_{b)c}=0,
\end{equation}
where the indices are raised and lowered with the background metric. The symbol $\textrm{Tr}(h)$ stands for $g^{ab}h_{ab}$, while the round brackets indicate a symmetrization with respect to the relevant indices, including the prefactor $\frac{1}{2}$.

Notice that, in this section, we will alternate between a notation where indices are explicit and one where they are implicit. This choice is related to our desire to avoid whenever possible a heavy notation where multiple subscripts appear.

 As much as Einstein's theory comes together with invariance under the action of the diffeomorphism group, so \eqref{lindyn} comes endowed with the linear counterpart. In other words two solutions $h,h^\prime$ of \eqref{lindyn} are said to be gauge equivalent if there exists $\chi\in\Gamma(T^*M)$ such that 
$$h^\prime = h+\nabla_S\chi,$$
where $\left(\nabla_S\chi\right)_{ab}\doteq\nabla_{(a}\chi_{b)}$. Per direct inspection one can realize that the operator $K$ in \eqref{lindyn} is not normally hyperbolic and thus one cannot construct smooth solutions using Green operators. Yet, one must keep in mind that, in gauge theories, we are not interested in single smooth solutions but actually in gauge equivalence classes of solutions. Hence, one can start from \eqref{lindyn} and look for a gauge transformation which reduces the dynamics to that ruled by a normally hyperbolic operator or by one which at least admits Green operators. The following proposition shows that it is indeed possible -- see for example \cite{Fewster:2012bj}:

\begin{prop}
Let 
$$\mathcal{S}_K(M)=\left\{h\in\Gamma(S^2T^*M)\;|\;Kh=0\right\},$$ 
be the space of smooth solutions to \eqref{lindyn} and let 
$$\mathcal{G}(M)=\left\{h\in\Gamma(S^2T^*M)\;|\;\exists\chi\in\Gamma(T^*M)\,\textrm{for which}\, h=\nabla_S\chi\right\}.$$
Then, for every $[h]\in\frac{\mathcal{S}_K(M)}{\mathcal{G}}$, there exists a representative $\widetilde h$ such that 
\begin{equation}\label{gaugefix}
\left\{\begin{array}{l}
\widetilde P\widetilde h=\left(\Box-2Riem\right)(I\widetilde h)=0\\
\textrm{div}(I\widetilde h)=0
\end{array}\right.,
\end{equation}
where $Riem$ is the Riemann tensor built out of $g$, $I$ is the trace reversal operator such that $(I\widetilde h)_{ab}=\widetilde h_{ab}-\frac{g_{ab}}{2}\textrm{Tr}(\widetilde h)$ while $\textrm{div}$ is the divergence operator such that $(\textrm{div}(\widetilde h))_b=\nabla^a\widetilde h_{ab}$.
\end{prop}

In this last proposition we have introduced the standard de Donder gauge and it is noteworthy since $\widetilde P$ is manifestly the composition of a normally hyperbolic operator with a trace reversal. Hence, adapting Definitions \ref{tc} and \ref{Green} to the case at hand, we associate to $\widetilde P$ the advanced (-) and the retarded (+) fundamental solutions $G^\pm_{\widetilde P}\doteq G^\pm_{\Box-2Riem}\circ I:\Gamma_{tc}(S^2T^*M)\to\Gamma(S^2T^*M)$ which enjoy the properties that, for every $\beta\in\Gamma_{tc}(S^2T^*M)$
\begin{gather*}
(\widetilde P\circ G^\pm_{\widetilde{P}})(\beta)=\beta=(G^\pm_{\widetilde{P}}\circ\widetilde P)(\beta),\\
\textrm{supp}(G^\pm_{\widetilde{P}}(\beta))\subseteq J^\pm(\textrm{supp}(\beta)).
\end{gather*}
Notice that $G^\pm_{\Box-2Riem}$ stands for the advanced/retarded fundamental solution of the normally hyperbolic operator $\Box-2Riem$.
Additionally we call $G_{\widetilde P}=G^-_{\widetilde P}-G^+_{\widetilde P}$ the {\em causal propagator of $\widetilde P$}. Yet, contrary to the scalar case, we cannot use only $G_{\widetilde P}$ to characterize in a covariant way the space of solutions of linearized gravity since we need also to take into account two additional data. On the one hand there is the gauge-fixing condition $\textrm{div}(I\widetilde h)=0$. This can be implemented by suitably restricting the space of admissible initial data, or more appropriately the admissible smooth and timelike compact sections of $S^2T^*M$, exploiting that $\textrm{div}\circ I\circ G^\pm_{\widetilde P}=G^\pm_\Box\circ\textrm{div}$, where $G^\pm_\Box$ are the advanced (-) and the retarded (+) Green operators for the d'Alembert wave operator acting on $\Gamma(T^*M)$. On the other hand \eqref{gaugefix} is not a complete gauge fixing. As a matter of fact, if $\widetilde h$ is a solution of \eqref{gaugefix}, so is every $\widetilde h^\prime$ such that $\widetilde h^\prime-\widetilde h=\nabla_S\chi$, $\chi\in\Gamma(T^*M)$ and $\Box\chi=0$. Adding together these two additional data, one can prove the following \cite{Benini:2014rya, Fewster:2012bj} 

\begin{prop}
There exists an isomorphism between $\frac{\mathcal{S}_K(M)}{\mathcal{G}(M)}$ and $\frac{Ker_{tc}(\textrm{div})}{Im_{tc}(K)}$ where 
$Ker_{tc}(\textrm{div})=\{\beta\in\Gamma_{tc}(S^2T^*M)\;|\;\textrm{div}(\beta)=0\}$ and $Im_{tc}(K)=\{\beta\in\Gamma_{tc}(S^2T^*M)\;|\;\beta=K(\beta^\prime),\;\beta^\prime\in\Gamma_{tc}(S^2T^*M)\}$. The isomorphism is realized by the causal propagator via $[\beta]\in \frac{Ker_{tc}(\textrm{div})}{Im_{tc}(K)}\mapsto [G(\beta)]\in\frac{\mathcal{S}_K(M)}{\mathcal{G}(M)}$.
\end{prop}

Having under control the space of dynamical configurations for linearized gravity we can proceed to introducing classical linear observables along the same lines as in Section \ref{2}. There are two important differences which we shall point out. First of all we start from all {\em kinematical configurations} $\Gamma(S^2T^*M)$ and, for every $\epsilon\in\Gamma_0(S^2TM)$, we define the linear functional
\begin{equation}\label{kinobs}
\mathcal{O}_\epsilon:\Gamma(S^2T^*M)\to\mathbb{R},\quad h\mapsto \mathcal{O}_\epsilon(h)=(\epsilon,h)\doteq\int\limits_M d\mu_g \epsilon^{ab}h_{ab}.
\end{equation}
The map $\mathcal{O}_\epsilon$ plays the role of a classical observable for kinematic configurations. At this stage comes the first difference from the scalar case, namely we need to encode the information of gauge invariance. This can be done by restricting our attention to those functionals of the form \eqref{kinobs} which vanish on all pure gauge configurations, that is those $h\in\Gamma(S^2T^*M)$ such that $h=\nabla_S\chi$, $\chi\in\Gamma(T^*M)$. Using \eqref{kinobs} and integration by parts, this entails
$$\mathcal{O}_\epsilon(\nabla_S\chi)=(\epsilon,\nabla_S\chi)=(\textrm{div}\epsilon,\chi)=0.$$
The arbitrariness of $\chi$ and the non-degenerateness of the pairing $(,)$ entails that $\textrm{div}\epsilon=0$. In other words, we can introduce the space of gauge invariant functionals 
$$\mathcal{L}^{inv}(M)=\{\mathcal{O}_\epsilon:\Gamma(S^2T^*M)\to\mathbb{R},\;|\;\textrm{div}\epsilon=0,\,\epsilon\in\Gamma_0(S^2TM)\}.$$
As last step, we need to account for dynamics which can be done by restricting the domain of definition of the observables from $\Gamma(S^2T^*M)$ to $\mathcal{S}_K(M)$. Identifying once more the space of linear observables with its labeling space, hence $\mathcal{L}^{inv}(M)$ with $Ker_0(\textrm{div})=\{\epsilon\in\Gamma_0(S^2TM)\;|\;\textrm{div}(\epsilon)=0\}$, the above restriction entails that $Ker_0(\textrm{div})$ includes redundant observables which we need to quotient. As shown in \cite{Benini:2014rya} this is tantamount to defining the following as the space of {\em classical observables for linearized gravity}:
\begin{equation}\label{dynobs}
\mathcal{L}^{obs}(M)\doteq\frac{\mathcal{L}^{inv}(M)}{Im_0(K)},
\end{equation}
where $Im_0(K)=\{\epsilon\in\Gamma_0(S^2TM)\;|\;\epsilon=K(\alpha),\;\alpha\in\Gamma_0(S^2TM)\}$. Notice the second big difference from the scalar case. While $\mathcal{E}^{obs}(M)$ could be endowed with a symplectic form, this is not necessarily the case for $\mathcal{L}^{obs}(M)$, to which we can associate the following pre-symplectic form:
\begin{equation}\label{presymp}
\tau:\mathcal{L}^{obs}(M)\times\mathcal{L}^{obs}(M)\to\mathbb{R},\quad([\epsilon],[\epsilon^\prime])\mapsto(\epsilon, G_{\widetilde P}(\epsilon^{\prime\flat}),
\end{equation}
where $G_{\widetilde P}$ is the causal propagator of $\widetilde P$, while $^\flat$ is the canonical, metric-induced musical isomorphism. It is important to stress that it is known that $\tau$ is non-degenerate if the Cauchy surface of $M$ is compact -- see for example \cite{Fewster:2012bj} -- or on Minkowski spacetime -- see \cite{Hack}. To the best of our knowledge, in all other cases the problem is still open. 

Having chosen a space of classical observables entails that we can define a $*$-algebra of observables associated to the quantum theory.

\begin{defn}\label{F_g(M)}
We call {\bf algebra of fields} for linearized gravity, the quotient $\mathcal{F}_{grav}(M)=\frac{\mathcal{T}_{grav}(M)}{\mathcal{I}_{grav}(M)}$. Here 
$$\mathcal{T}_{grav}(M)\doteq \bigoplus_{n=0}^\infty\mathcal{L}^{obs}(M;\mathbb{C})^{\otimes n},$$
is the universal tensor algebra endowed with complex conjugation as $*$-operation, where $\mathcal{L}^{obs}(M;\mathbb{C})=\mathcal{L}^{obs}(M)\otimes\mathbb{C}$ while $\mathcal{L}^{obs}(M;\mathbb{C})^{\otimes 0}\doteq\mathbb{C}$. $\mathcal{I}_{grav}(M)$ is the $*$-ideal generated by the {\em canonical commutation relations} $[\epsilon]\otimes [\epsilon^\prime]-[\epsilon^\prime]\otimes [\epsilon]-i\tau\left([\epsilon],[\epsilon^\prime]\right)\mathbb{I}$, where $\mathbb{I}$ is the identity in $\mathcal{T}_{grav}(M)$ and $\tau$ is defined in \eqref{presymp}.
\end{defn}

Notice that, since $\tau$ is not known to be symplectic, we cannot conclude that $\mathcal{F}_{grav}(M)$ is semisimple. Hence it might contain an Abelian ideal, namely there might exists observables behaving classically regardless of the state chosen for $\mathcal{F}_{grav}(M)$. Furthermore one can also show that $\mathcal{F}_{grav}(M)$ satisfies the counterpart of Proposition \ref{properties} for $\mathcal{F}(M)$: causality and the time-slice axiom. We will not enter into the details to avoid useless repetitions. The next step in our construction will be the identification of algebraic states of physical interest for linearized gravity. While the definition of an algebraic state and the content of Theorem \ref{GNS}
are left unchanged, the Hadamard condition requires to be slightly adapted to account for gauge invariance. More precisely, since we will be mainly interested in Gaussian states for $\mathcal{F}_{grav}(M)$, we will be looking for two-point functions $\omega_2:\mathcal{L}^{obs}(M)\otimes\mathcal{L}^{obs}(M)\to\mathbb{R}$. In view of \eqref{dynobs}, this is tantamount to building $\Omega_2:\mathcal{L}^{inv}(M)\otimes\mathcal{L}^{inv}(M)\to\mathbb{R}$, weak bi-solution of \eqref{lindyn}. Yet, since, on account of gauge invariance, $\mathcal{L}^{inv}(M)$ includes only those $\epsilon\in\Gamma_0(S^2TM)$ which are divergence free, we cannot ensure automatically that $\Omega_2$ identifies a bi-distribution $\widetilde\Omega_2:\Gamma_0(S^2TM)\times\Gamma_0(S^2TM)\to\mathbb{R}$. Additionally we need to take into account that, besides $\widetilde P$, also the trace-reversal is present in \eqref{lindyn}. Following \cite{Fewster:2012bj}, we define a trace-reversal operation at the level of of bi-distributions and, with a slight abuse of notation, we indicate it still with the letter $I$. Let thus $\widetilde\Omega_2$ be a bi-distribution on $\Gamma_0(S^2TM)$ we call {\em trace reversal} of $\widetilde\Omega_2$
\begin{gather*}
I\widetilde\Omega_2:\Gamma_0(S^2TM)\times\Gamma_0(S^2TM)\to\mathbb{R},\\
(\epsilon,\epsilon^\prime)\mapsto (I\widetilde\Omega_2)(\epsilon,\epsilon^\prime)=\widetilde\Omega_2(\epsilon,\epsilon^\prime)-\frac{1}{8}\textrm{Tr}(\widetilde\Omega_2)(\textrm{Tr}(\epsilon),\textrm{Tr}(\epsilon^\prime)),
\end{gather*}
where $\textrm{Tr}(\widetilde\Omega_2)\in\mathcal{D}^\prime(M\times M)$ is defined as follows: for all $f,f^\prime\in C^\infty_0(M)$,
$$\textrm{Tr}(\widetilde\Omega_2)(f,f^\prime)\doteq\widetilde\Omega_2(g^{-1}f, g^{-1}f^\prime),$$
$g^{-1}$ being the inverse metric. To summarize we define \cite{Fewster:2012bj}
\begin{defn}\label{Hadgrav}
Let $\omega:\mathcal{F}_{grav}(M)\to\mathbb{C}$ be a quasi-free state. It is said to be {\bf Hadamard} if there exists a bi-distribution $\widetilde\Omega_2:\Gamma_0(S^2TM)\times\Gamma_0(S^2TM)\to\mathbb{R}$ which is a weak bi-solution of $\widetilde P$, its wavefront set has the same form of \eqref{WF} and, for every $\epsilon,\epsilon^\prime\in\mathcal{L}^{inv}(M)$
$$\omega([\epsilon]\otimes[\epsilon^\prime])=(I\widetilde\Omega_2)(\epsilon,\epsilon^\prime).$$
\end{defn}

\subsection{Hadamard states from null infinity - II}\label{5}
In this section we will show how the bulk-to-boundary correspondence can be applied to linearized gravity. Hence, from now on $(M,g)$ will indicate a globally hyperbolic, asymptotically flat spacetime with vanishing Ricci tensor and $(\widetilde M,\widetilde g)$ the associated unphysical spacetime. The most notable difference between a massless, conformally coupled real scalar field and linearized gravity consists of the behaviour under a conformal transformation of the equations ruling the dynamics. As a matter of fact, by mapping $g$ to $\Xi^2 g$, \eqref{lindyn} transforms in such a way that several terms proportional to inverse powers of $\Xi$ appear. Taking into account that the null boundary of an asymptotically flat spacetime is the locus $\Xi=0$, such behaviour is clearly problematic. This feature is not proper only of linearized gravity, but also of free electromagnetism, written in terms of the vector potential. The solution in all these cases is to exploit gauge invariance, namely to look for a suitable gauge fixing which makes the dynamics hyperbolic and, upon a conformal transformation, controls all possible divergences due to the terms proportional to inverse powers of $\Xi$. While, in free electromagnetism, the standard Lorenz gauge is the right choice, for linearized gravity, the de Donder gauge is not suitable for this task. The problem was tackled in the literature, especially in connection to the stability of asymptotically simple spacetimes and an answer was found going under the name of {\em Geroch-Xanthopoulos} gauge \cite{GX}.

Let us now go into the details of the construction. As in the scalar case, the starting point is the identification of a suitable space of tensors living on future null infinity. Following \cite{Ashtekar:1982aa}, we define
$$\widetilde{\mathcal{S}}(\mathscr{I}^+)=\{\lambda\in\Gamma(S^2T^*\mathscr{I}^+)\;|\;\lambda_{ab}n^a=0\;\textrm{and}\;\lambda_{ab}q^{ab}=0\},$$
where $n_a=\widetilde\nabla_a\Xi$ and $\widetilde\nabla$ is the covariant derivative built out of $\widetilde g$. The tensor $q=\iota^*g$ where $\iota:\mathscr{I}^+\to\widetilde M$ and $q^{ab}$ is any inverse such that $q^{ab}q_{ac}q_{bd}=q_{cd}$. Subsequently we consider a vector subspace 
$$\mathcal{S}_{grav}(\mathscr{I}^+)\doteq\{\lambda\in\widetilde{\mathcal{S}}(\mathscr{I}^+)\;|\;(\lambda,\lambda)_\mathscr{I}<\infty,\quad\textrm{and}\quad(\partial_u\lambda,\partial_u\lambda)_\mathscr{I}<\infty\},$$
where, for any $\lambda,\lambda^\prime\in\widetilde{\mathcal{S}}(\mathscr{I}^+)$,
$$(\lambda,\lambda^\prime)_\mathscr{I}\doteq\int\limits_\mathscr{I}d\mu_\mathscr{I} \lambda_{ab}\lambda^\prime_{cd}q^{ac}q^{bd}.$$
As for the real scalar case, the space of functions on future null infinity has been chosen since it enjoys two important properties:
\begin{enumerate}
\item It is a symplectic space if endowed with the following antisymmetric bilinear form:
\begin{gather}
\tau_\mathscr{I}:\mathcal{S}_{grav}(\mathscr{I}^+)\times\mathcal{S}_{grav}(\mathscr{I}^+)\to\mathbb{R},\notag\\
(\lambda,\lambda^\prime)\mapsto\tau_\mathscr{I}(\lambda,\lambda^\prime)=\int\limits_\mathscr{I}d\mu_\mathscr{I}(\lambda_{ab}\mathcal{L}_n\lambda^\prime_{cd}-\lambda^\prime_{ab}\mathcal{L}_n\lambda_{cd})q^{ac}q^{bd},\label{sympl2}
\end{gather}
where $\mathcal{L}_n$ is the Lie derivative along the vector field $n$ on $\mathscr{I}^+$. 
\item the pair $(\mathcal{S}_{grav}(\mathscr{I}^+),\tau_\mathscr{I})$ is invariant under the following representation $\Pi$ of the BMS group \eqref{BMS}: Let us fix a Bondi frame on $\mathscr{I}^+$, let $(\Lambda,\alpha(z,\bar{z}))\in\textrm{BMS}$ and let $\lambda\in\mathcal{S}_{grav}(\mathscr{I}^+)$; then, recalling \eqref{KLambda},
\begin{equation}\label{Pi}
\left[\Pi_{(\Lambda,\alpha)}\lambda\right](u,z,\bar{z})=K_\Lambda(z,\bar{z})\lambda(u+\alpha(z,\bar{z}),z,\bar{z}).
\end{equation}
\end{enumerate}
In view of these properties of the space of functions, which we use on future null infinity, we can define an auxiliary $*$-algebra on $\mathscr{I}^+$:
\begin{defn}\label{Scrigrav}
We call {\em $*$-algebra of fields for linearized gravity on future null infinity}, $\mathcal{F}_{grav}(\mathscr{I}^+)=\frac{\mathcal{T}_{grav}(\mathscr{I}^+)}{\mathcal{I}_{grav}(\mathscr{I}^+)}$, where  
$$\mathcal{T}_{grav}(\mathscr{I}^+)\doteq\bigoplus_{n=0}^\infty\mathcal{S}_{grav}(\mathscr{I}^+;\mathbb{C})^{\otimes n}.$$
Here $\mathcal{S}_{grav}(\mathscr{I}^+;\mathbb{C})=\mathcal{S}_{grav}(\mathscr{I}^+)\otimes\mathbb{C}$ whereas $\mathcal{S}_{grav}(\mathscr{I}^+;\mathbb{C})^{\otimes 0}\doteq\mathbb{C}$. At the same time $\mathcal{I}_{grav}(\mathscr{I}^+)$ is the $*$-ideal generated by the relation $\lambda\otimes\lambda^\prime-\lambda^\prime\otimes\lambda-i\tau_{\mathscr{I}}(\lambda,\lambda^\prime)\mathbb{I}$, where $\mathbb{I}$ is the identity in $\mathcal{T}_{grav}(\mathscr{I}^+)$. The $*$-operation is complex conjugation.
\end{defn} 

The next step consists of associating to each classical linear observable in $(M,g)$ an element of $\mathcal{S}_{grav}(\mathscr{I}^+)$. We follow again the procedure devised in \cite{Ashtekar:1982aa}. First of all we remark that, to every $[\epsilon]\in\mathcal{L}^{obs}(M)$, we can associate $E_{\widetilde P}(\epsilon^\flat)$, which is a smooth and spacelike compact solution of \eqref{gaugefix} and thus also of \eqref{lindyn}. 

\begin{defn}\label{radobs}
We call {\em classical radiative observables}, $\mathcal{L}^{rad}(M)$ the collection of all $[\epsilon]\in\mathcal{L}^{obs}(M)$, for which $E_{\widetilde P}(\epsilon^\flat)$ is gauge equivalent to a smooth and spacelike compact solution $h$ of \eqref{lindyn} in the {\bf Geroch-Xanthopoulos gauge}, that is, setting $\gamma_{ab}=\Xi \gamma^\prime_{ab}$, $\gamma_a=\Xi^{-1} n^b \gamma^\prime_{ab}$, $\gamma=\widetilde g^{ab}\gamma_{ab}$ and $f=\Xi^{-1}n^a n_a$ and $n_a=\nabla_a\Xi=\widetilde\nabla_a\Xi$, it holds that
\begin{subequations}
\begin{align}
y_a=\widetilde\nabla^b\gamma_{ab}-\widetilde\nabla_a\gamma-3\gamma_a & =0,\label{GXa}\\
\left(n^a \widetilde\nabla_a+\frac{1}{6}\Xi \widetilde R+\frac{3}{2}f\right)\widetilde\Box\gamma & = \frac{1}{12}\widetilde R f\gamma - \frac{1}{2}\gamma\widetilde\Box f-\frac{1}{3}\widetilde R n^a\gamma_a+\frac{4}{\Xi} \widetilde C_{abcd}\gamma^{bd}n^an^c , \label{GXb}
\end{align}
\end{subequations}
where $\widetilde\cdot$ refers to quantities computed with respect to $\widetilde g$, {\em e.g.}, $\widetilde C_{abcd}$ is the Weyl tensor for $\widetilde g$. At the same time we call {\em algebra of radiative observables} $\mathcal{F}^{rad}(M)$ the $*$-subalgebra of $\mathcal{F}_{grav}(M)$ built of $\mathcal{L}^{rad}(M)$.
\end{defn}

Radiative observables play a distinguished role since, as shown in \cite{Ashtekar:1982aa, GX}, \eqref{lindyn} in combination with the Geroch-Xanthopoulos gauge yields an hyperbolic system of partial differential equations on $(\widetilde M, \widetilde g)$. Thus every $\gamma=\Xi h$, $h=E_{\widetilde P}(\epsilon^\flat)$ and $[\epsilon]\in\mathcal{L}^{rad}(M)$, can be uniquely extended to a smooth solution of the conformally transformed equations of motion for linearized gravity in the unphysical spacetime. More importantly such extension can be restricted to future null infinity and, as proven in \cite{Ashtekar:1982aa, Benini:2014rya},

\begin{prop}\label{b2b}
Let us endow $\mathcal{L}^{rad}(M)$ with the restriction of \eqref{presymp} thereon. Then, there exists a map $\Upsilon_{grav}:\mathcal{E}^{rad}(M)\to\mathcal{S}_{grav}(\mathscr{I}^+)$ such that, for all $[\epsilon],[\epsilon^\prime]\in\mathcal{E}^{rad}(M)$
$$\tau_\mathscr{I}([\epsilon],[\epsilon^\prime])=\tau(\Upsilon_{grav}([\epsilon]),\Upsilon_{grav}([\epsilon^\prime])),$$
where $\tau_\mathscr{I}$ is defined in \eqref{sympl2}. $\Upsilon_{grav}$ can be extended to a $*$-homomorphism $\iota_{grav}:\mathcal{F}^{rad}(M)\to\mathcal{F}_{grav}(\mathscr{I}^+)$ which is completely defined by its action on the generators, namely, for every $[\epsilon]\in\mathcal{L}^{rad}(M)$, $\iota_{grav}([\epsilon])\doteq\Upsilon_{grav}([\epsilon])$.
\end{prop}

Notice that, contrary to the scalar case and due to our lack of control on the non-degenerateness of \eqref{presymp}, we cannot conclude that $\iota_{grav}$ is injective. Yet this is no obstacle for constructing states on $\mathcal{F}^{rad}(M)$ via a pull-back of those for $\mathcal{F}_{grav}(\mathscr{I}^+)$. Before investigating this problem, we need to answer an important question, namely if $\mathcal{L}^{rad}(M)$ coincides with $\mathcal{L}^{obs}(M)$. In the original paper \cite{GX}, it appeared as if every smooth and spacelike compact solution of \eqref{lindyn} could be transformed into one in the Geroch-Xanthopoulos gauge. Yet, a closer investigation of the procedure unveils the presence of obstructions. Most surprisingly it turns out that problems arise in implementing \eqref{GXa} rather than \eqref{GXb}. More precisely let $h\in\Gamma(S^2T^*M)$ be a solution of \eqref{lindyn} and let $h^\prime=h+\nabla_S\chi$, $\chi\in\Gamma(T^*M)$. Then, a direct computation shows that $h^\prime$ satisfied \eqref{GXa} if and only if 
$$\nabla^b\nabla_{[b}\chi_{a]}=-v_a(h),\quad v_a(h)\doteq \nabla^b h_{ab}-\nabla_a h,$$
where the square brackets between the subscripts stand for total antisymmetrization, including the prefactor $\frac{1}{2}$. This identity entails that $v\in\Gamma(T^*M)$ must be a co-exact $1$-form, a property which is not obviously enjoyed. In \cite{Benini:2014rya} it has been proven the following:
\begin{prop}\label{MainProp}
Let $h=E_{\widetilde{P}}(\epsilon^\flat)$ with $[\epsilon]\in\mathcal{L}^{obs}(M)$. Then $h$ is gauge equivalent to a solution of \eqref{lindyn} if and only if $\textrm{Tr}(\epsilon)=g_{ab}\epsilon^{ab}$ is the codifferential of a compactly supported $1$-form.
\end{prop}

Proposition \ref{MainProp} offers a more practical condition to verify the implementability of the Geroch-Xanthopoulos gauge, although the question is still non trivial since one has to account for two mixing conditions, $\epsilon$ being divergence free and its trace being co-exact. Yet, in \cite{Benini:2014rya} it has been proven that, while on Minkowski spacetime the hypotheses of Proposition \ref{MainProp} are met, there exist asymptotically flat, globally hyperbolic and Ricci flat spacetimes for which they are not. Most notably it suffices that the Cauchy surface $\Sigma$ of $(M,g)$ is diffeomorphic to $X\times\mathbb{S}^1$, $X$ being a codimension $1$ submanifold of $\Sigma$ and that that $g$ admits a Killing field along $\mathbb{S}^1$. 

Having understood under which conditions radiative observables coincides with all classical linear ones, we can revert to our main investigation, namely the construction of Hadamard states. At this stage the procedure will be identical to the one described in Section \ref{3} and we shall only sketch the key points. The starting one is Proposition \ref{b2b} and in particular the $*$-homomorphism $\iota_{grav}$. First of all we identify a distinguished quasi-free state for $\mathcal{F}_{grav}(\mathscr{I}^+)$:
\begin{prop}\label{2pt}
The map $\omega_2^\Im:\mathcal{S}_{grav}(\mathscr{I}^+;\mathbb{C})\otimes\mathcal{S}(\mathscr{I}^+;\mathbb{C})\to\mathbb{R}$ such that 
\begin{equation}\label{eq2pt}
\omega_2^\Im(\lambda\otimes \lambda^\prime)=-\frac{1}{\pi}\lim_{\epsilon\to 0}\int\limits_{\mathbb{R}^2\times\mathbb{S}^2}\frac{\lambda_{ab}(u,\theta,\varphi)\lambda^{\prime}_{cd}(u^\prime,\theta,\varphi)q^{ac}q^{bd}}{(u-u^\prime-i\epsilon)^2}\,d u\,d u^\prime\,d\mathbb{S}^2(\theta,\varphi),
\end{equation}
where $d\mathbb{S}^2(\theta,\varphi)$ is the standard line element on the unit $2$-sphere, unambiguously defines a quasi-free state $\omega^\Im:\mathcal{F}_{grav}(\mathscr{I}^+)\to\mathbb{C}$. Furthermore:
\begin{enumerate}
\item $\omega^\Im$ induces via pull-back a quasi-free bulk state $\omega^M:\mathcal{F}^{rad}(M)\to\mathbb{C}$ such that $\omega^M\doteq\omega^\Im\circ\Upsilon$,
\item $\omega^\Im$ is invariant under the action $\Pi$ of the BMS group induced by \eqref{Pi} on $\mathcal{F}_{grav}(\mathscr{I}^+)$.
\end{enumerate}
\end{prop}
Notice that we use the symbol $\Pi$ with a slight abuse of notation since we have already introduced it to indicate in \eqref{Pi} the representation of the BMS group on $\mathcal{S}(\mathscr{I}^+)$. Since $\mathcal{F}_{grav}(\mathscr{I}^+)$ is built out of $\mathcal{S}(\mathscr{I}^+)$ we feel that no confusion can arise. As a last step we need to combine \eqref{eq2pt} with Definition \ref{Hadgrav} to conclude that, for all those asymptotically flat and globally hyperbolic spacetimes for which $\mathcal{E}^{obs}(M)=\mathcal{E}^{rad}(M)$, it holds -- see \cite{Benini:2014rya} for the proof:
\begin{thm}
Let $\iota:\mathcal{F}^{rad}(M)\to\mathcal{F}_{grav}(\mathscr{I}^+)$ be as in Proposition \ref{b2b}. The state $\omega^M=\omega^\Im\circ\iota:\mathcal{F}^{rad}(M)\to\mathbb{C}$, where $\omega^\Im$ is the state introduced in Proposition \ref{2pt}, enjoys the following properties:
\begin{enumerate}
\item Its two-point function is the restriction to $\mathcal{E}^{rad}(M)\times\mathcal{E}^{rad}(M)$ of a bi-distribution on $\widetilde M$ whose wavefront set on $M$, seen as an open submanifold of $\widetilde M$, is of Hadamard from,
\item It is invariant under the action of all isometries of the bulk metric $g$, that is $\omega^M\circ\alpha_\phi=\omega^M$. Here $\phi:M\to M$ is any isometry and $\alpha_\phi$ represents the action of $\phi$ induced on $\mathcal{F}^{rad}(M)$ by setting $\alpha_\phi([\epsilon])=[\phi_*\epsilon]$ on the algebra generators $[\epsilon]\in\mathcal{E}^{rad}(M)$;
\item It coincides with the Poincar\'e vacuum on Minkowski spacetime.
\end{enumerate}
\end{thm}


\subsection*{Acknowledgment}
The author is grateful to the University of Regensburg and to the organizing committee for the kind hospitality during the workshop {\em Quantum Mathematical Physics} held from the 29th of September to 2nd of October 2014. C. D. is grateful to Simone Murro for the useful discussions.

\end{document}